\documentclass[aps,dvipsnames,twocolumn,floatfix,amsmath]{revtex4-1}
\usepackage{graphicx}%
\relax

\begin{document}

\title{Evidence for room temperature superconductivity at graphite interfaces
}


\author{Pablo D. Esquinazi}
\affiliation{Division of Superconductivity and Magnetism, Felix
Bloch Institute for Solid State Physics, Universit\"{a}t Leipzig,
Linn\'{e}stra{\ss}e 5, D-04103 Leipzig, Germany}
\author{Christian E. Precker}
\affiliation{Division of Superconductivity and Magnetism, Felix
Bloch Institute for Solid State Physics, Universit\"{a}t Leipzig,
Linn\'{e}stra{\ss}e 5, D-04103 Leipzig, Germany}
\author{Markus Stiller}
\affiliation{Division of Superconductivity and Magnetism, Felix
Bloch Institute for Solid State Physics, Universit\"{a}t Leipzig,
Linn\'{e}stra{\ss}e 5, D-04103 Leipzig, Germany}
\author{Tiago R. S. Cordeiro}
\affiliation{Division of Superconductivity and Magnetism, Felix
Bloch Institute for Solid State Physics, Universit\"{a}t Leipzig,
Linn\'{e}stra{\ss}e 5, D-04103 Leipzig, Germany}
\author{Jos\'e Barzola-Quiquia}
\affiliation{Division of Superconductivity and Magnetism, Felix
Bloch Institute for Solid State Physics, Universit\"{a}t Leipzig,
Linn\'{e}stra{\ss}e 5, D-04103 Leipzig, Germany}
\author{Annette Setzer}
\affiliation{Division of Superconductivity and Magnetism, Felix
Bloch Institute for Solid State Physics, Universit\"{a}t Leipzig,
Linn\'{e}stra{\ss}e 5, D-04103 Leipzig, Germany}
\author{Winfried B\"ohlmann }
\affiliation{Division of Superconductivity and Magnetism, Felix
Bloch Institute for Solid State Physics, Universit\"{a}t Leipzig,
Linn\'{e}stra{\ss}e 5, D-04103 Leipzig, Germany}

\begin{abstract}
In the last 43 years several hints were reported suggesting the
existence of granular superconductivity above room temperature in
different graphite-based systems. In this paper some of the
results are reviewed, giving special attention to those obtained
in water and n-heptane treated graphite powders, commercial and
natural bulk graphite samples with different characteristics as
well as transmission electron microscope (TEM) lamellae. The
overall results indicate that superconducting regions exist and
are localized at certain internal interfaces of the graphite
structure. The existence of the rhombohedral graphite phase in all
samples with superconducting-like properties suggests its
interfaces with the Bernal phase as a possible origin for the
high-temperature superconductivity, as theoretical calculations
predict.  High precision  electrical resistance and magnetization
measurements were used to identify a transition at $T_c  \gtrsim
350~$K. To check for the existence of true zero resistance paths
in the samples we used local magnetic measurements, which results
support the existence of superconducting regions  at such high
temperatures.

\end{abstract}

\maketitle
\bigskip

\section{First hints for room temperature superconductivity in disordered graphite powders}
\label{intro}

\begin{figure}
\begin{center}
\includegraphics[width=0.5\textwidth]{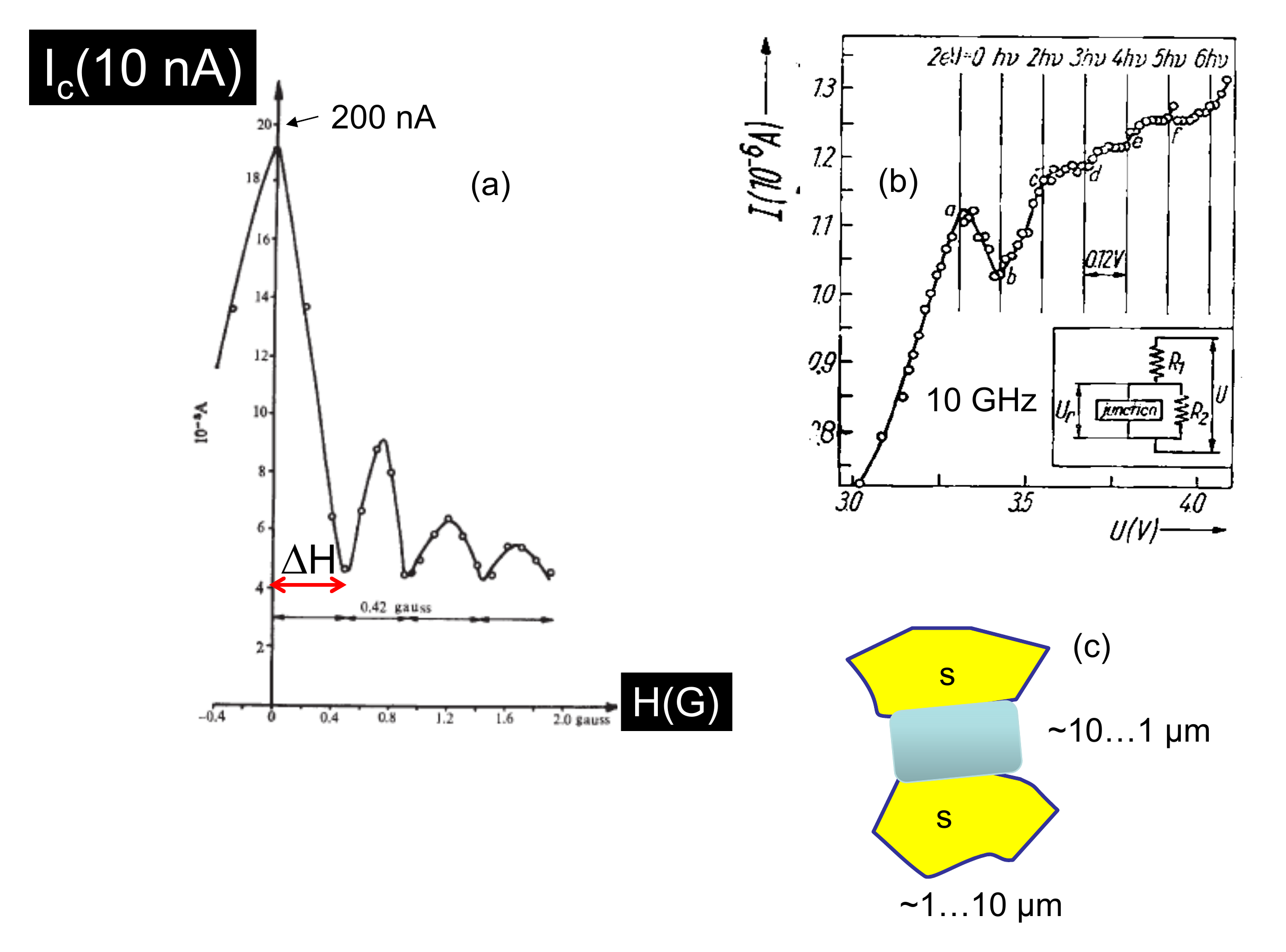}
\caption{(a) Maximum critical current vs. applied magnetic field
deduced from current-voltages curves obtained from annealed carbon
powder under an input current through  two Al electrodes at
room temperature. Upon sample, the field value at the first minimum
is $\Delta H = 0.1 \ldots 1~$G. Adapted from \cite{ant74}.
(b) Current-voltage characteristic line obtained at room temperature
for a similar annealed graphite powder as in (a) under a radiation
of 10~GHz frequency. The inset shows the effective electric
circuit proposed to explain the large voltage values between the
apparent steps. The values of the resistance $R_1,R_2$ are unknown. Adapted
from \cite{ant75}. (c) Sketch of a Josephson junction with the
superconducting parts (yellow boxes) and the (blue) region representing the tunnel barrier
plus the area where the magnetic flux
produced by an applied magnetic field applied normal to the area
 enters. This (blue) area would be given by $(2 \lambda
+ d)L$, which would be $ \sim 10 \ldots 100~\mu$m$^2$,  according to the range of
fields where the first minimum was measured  \cite{ant74}, see (a).}
\label{fig1}       
\end{center}
\end{figure}

In the year 1974 Kazimierz Antonowicz published studies
\cite{ant74} on current-voltage $I-V$ characteristics curves of
annealed carbon powder contacted between two Al electrodes. He
recognized a kind of ``critical current"  $I_c$ from the   $I-V$
curves and measured the influence of  low amplitude applied
magnetic fields on $I_c$. The result is reproduced in
Fig.~\ref{fig1}(a). The field dependence of this ``critical
current"  $I_c$ follows the field dependence given by the
Josephson critical current equation $I_c = |I_0  \sin(x)/x + I_1|$
whereas $x \propto H$, $H$ the applied field  and $I_1$ a
field-independent constant background current. This constant as
well as the field $\Delta H$ at which $I_c$ shows the first
minimum are sample dependent, the last between 0.1~Oe $< \Delta H
< 1$~Oe \cite{ant74}.

The observed behavior in Fig.~\ref{fig1}(a) follows the maximum
net  supercurrent equation for a Josephson junction under a
magnetic field in the plane of the junction, referred to as
Fraunhofer diffraction pattern, see e.g. \cite{tin96}. There are
several details we should note before arguing against or for a
possible interpretation on the basis of a Josephson
superconducting junction. Namely,  the way the sample and the
contacts were prepared did not allow for a direct contact to the
possible superconducting junctions, which may explain the finite
background $I_1$. If we assume that the observed behavior is
indeed due to a Josephson junction the field variable $x = \pi
\Phi/ \Phi_0$, where the enclosed magnetic flux for a rectangular
junction of thickness $d$ and length $L$ (normal to the applied
field), see Fig.~\ref{fig1}(c), would be $\Phi = (2 \lambda + d)L
H\mu_0$ with $\lambda$ the London penetration depth. The range of
field $\Delta H$ where the first minimum was observed would
indicate a rather large area $\sim 10~\mu$m$^2 < (2 \lambda + d)L
< ~\sim100~\mu$m$^2$. Assuming that there are superconducting
regions coupled through a non-superconducting path in the annealed
graphite powder, how could be possible that Cooper pairs tunnel
through a rather large carbon matrix? On the other hand, if
instead of a three dimensional a two dimensional (2D) junction is
formed, as we may have at certain interfaces (see
Section~\ref{inter}), taking into account that Cooper pairs can
survive large distances in a graphene layer \cite{hee07} and the
expected enhancement of the effective London penetration depth in
this 2D case \cite{pea64}, such a behavior does not appear
impossible.

The same author published one year later the behavior of the $I-V$
curves of the same annealed carbon powder at room temperature but
irradiated with radiation of 10~GHz frequency \cite{ant75}, see
Fig.~\ref{fig1}(b). For an ideal Josephson junction we would
expect that this radiation produces constant-voltage Shapiro steps
\cite{sha63} in the DC $I-V$ curves at voltages $V_n = n \hbar
\omega/2e$, with $n$ an integer and $\omega$ the angular frequency
of radiation. For $\omega = 2 \pi 10^{10}~$Hz we expect certain
steps in the $I-V$ curves at voltages $V_n = n 20~\mu$V, several
orders of magnitude smaller than the observed 120~mV, see
Fig.~\ref{fig1}(b). Following the author explanation for this
large voltage difference, one notes that the used experimental
setup
 did not test only the voltage coming from the apparent
Josephson junction but also the one provided by extra resistances,
see inset in Fig.~\ref{fig1}(b). The value of those resistance remained, however, unknown.

Independently of the uncertainties in the interpretation of the
experimental results shown in Fig.~\ref{fig1}, perhaps the main
obstacle we have to believe on the results reported in
\cite{ant74,ant75} is the fact that they were obtained at room
temperature. If we assume that the experiment and its
interpretation are correct, there are further details written
explicitly in those publications \cite{ant74,ant75} that one
should take into account if one wants to repeat  the
experiments, namely: The ``new supercondcuting state" is quasi-stable, detectable
for a few hours or it vanishes within a few days, and, only $\sim
30\%$ of the ``properly produced" samples showed the effect.
These hints  would indicate a superconductor very sensitive to the preparation details and rather unstable.
It is appealing to argue that the origin of the measured signals should be related to
certain interfaces between graphite-like  grains.

All these details clearly prevented a quick reproduction of the
published results added to the huge skepticism of the scientific
community. We are not aware of any published report showing
similar results as those  in \cite{ant74,ant75}. In what follows
we review some magnetization results obtained in water and
n-heptane treated high-quality graphite powders that suggest that
granular superconductivity might be indeed possible  in  the
powders, very probably at certain interfaces created after the
corresponding treatments \cite{sch12,schcar}.

\section{Magnetization measurements on graphite powders}
\label{powder}

In this section we discuss results obtained from graphite powders
that support to some extent  the results discussed in
Section~\ref{intro}. The magnetization results of graphite
powders, after a certain water or alkane treatment, indicate a
granular superconducting behavior that support the existence of
superconducting regions in graphite and, indirectly, the
possibility to have Josephson junctions in some regions.

\begin{figure}
  \includegraphics[width=0.5\textwidth]{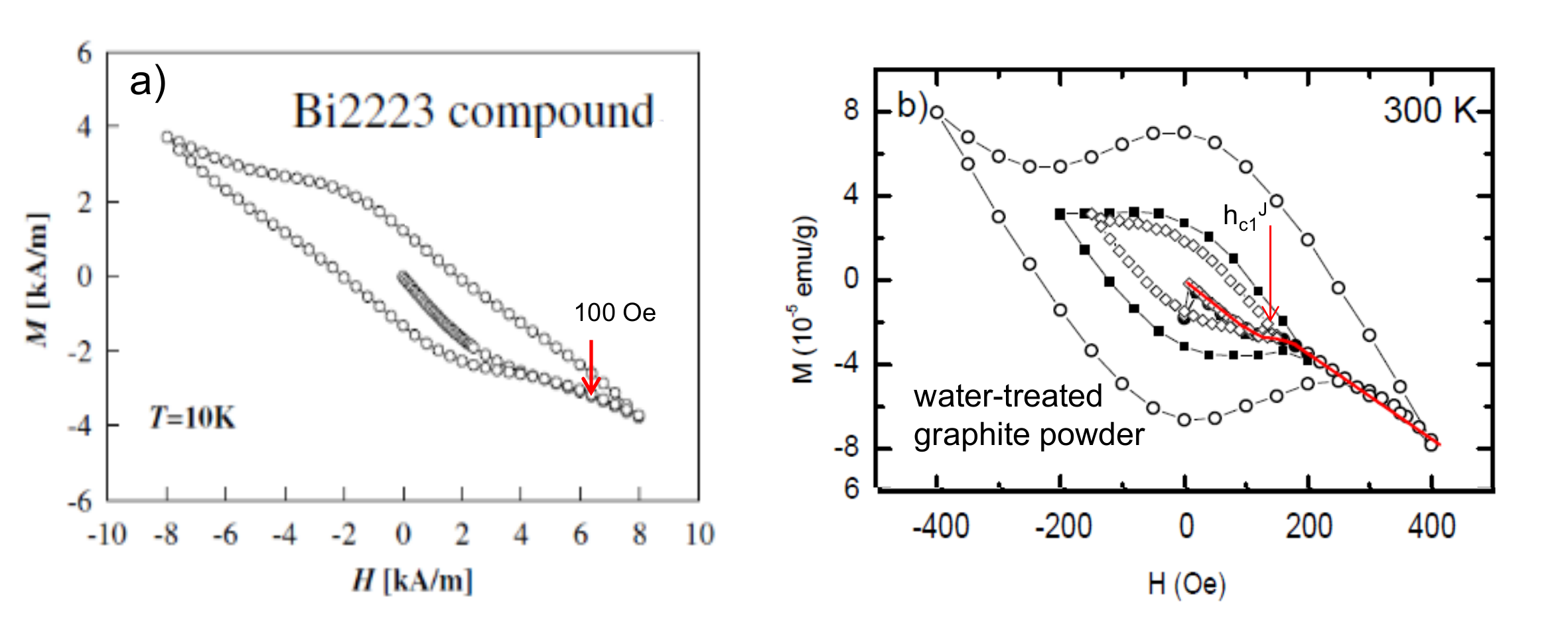}
\caption{a) Field dependence of the magnetization of a Bi2223
high-$T_c$ oxide granular compound at 10~K, adapted from \cite{and01}. b)
Similar plot for a water-treated graphite powder at different
maximum applied fields at 300~K, after subtraction of linear in field diamagnetic
background, adapted from \cite{sch12}.}
\label{fig4}       
\end{figure}

The behavior of the magnetization of granular superconductors has
been very well described in several works in the last 25 years,
see e.g. \cite{and01,sen91}. Due to the granularity and the
Josephson coupling between grains, the strength of an applied
magnetic field influences in a non simple way the field
hysteresis. For example, at fields lower than the first critical
field of the grains $H_{c1}$ and lower than the field $h_{c1}^J$
necessary to destroy the weakest links between the grains, a
granular  sample shows a Meissner-like response. At fields
above $h_{c1}^J$ but below the field $h_{c2}^J$ (where the
coupling between grains is completely overwhelmed and the
magnetization is imposed by the London currents circulating around
each of the grains) the hysteresis loop has a peculiar shape as
shown in Fig.~\ref{fig4}(a) \cite{sen91}. At fields above
$h_{c2}^J$ but below the intrinsic $H_{c1}$ of the grains, the
hysteresis vanishes up to the field $H_{c1}$, at which Abrikosov
vortices enters in the grains and field hysteresis loops are again
measured due to their finite pinning in the superconducting matrix
\cite{sen91}.

The magnetization of an untreated, very pure graphite powder, does
not show any remarkable hysteresis \cite{sch12}. However, after a
treatment  with water \cite{sch12} or n-heptane (following the
work in \cite{kaw13}) a clear irreversibility in field and in
temperature is observed. As example,  Fig.~\ref{fig4}(b) shows the
hysteresis loops of water-treated graphite powder at 300~K and at
different maximum applied fields, after subtraction of a linear
diamagnetic background. The hysteresis loops at fields below and
above $h_{c1}^J$ are qualitatively similar to the one obtained in
granular superconductors, see Figure~\ref{fig4}(a) and also
\cite{and01,sen91}. Similar curves were obtained after treatment
the graphite powder with alkanes. The field hysteresis loops, the
hysteresis in temperature, i.e. difference between the
magnetization at field cooled (FC) and zero field cooled (ZFC)
\cite{sch12}, suggest the existence of granular superconductivity
at certain regions formed after the liquid treatment. Because a
similar behavior is observed in graphite samples with interfaces
\cite{schcar}, it is appealing to suggest that certain interfaces
are produced after the liquid treatment of the graphite grains. We
note that the measurements were always done after the
corresponding liquid was evaporated. An important experimental
fact should be also stressed: If we apply pressure to the treated
powder, the hysteresis loop vanishes, indicating that the
hysteresis does not appear to be due to defect-induced magnetic
order but it comes from   certain regions of the graphite grains
that are destroyed with pressure \cite{sch12}.

\subsection{Thermomagnetic hysteresis measurements help to differentiate between a superconducting and ferromagnetic
behavior}

\begin{figure}
  \includegraphics[width=0.5\textwidth]{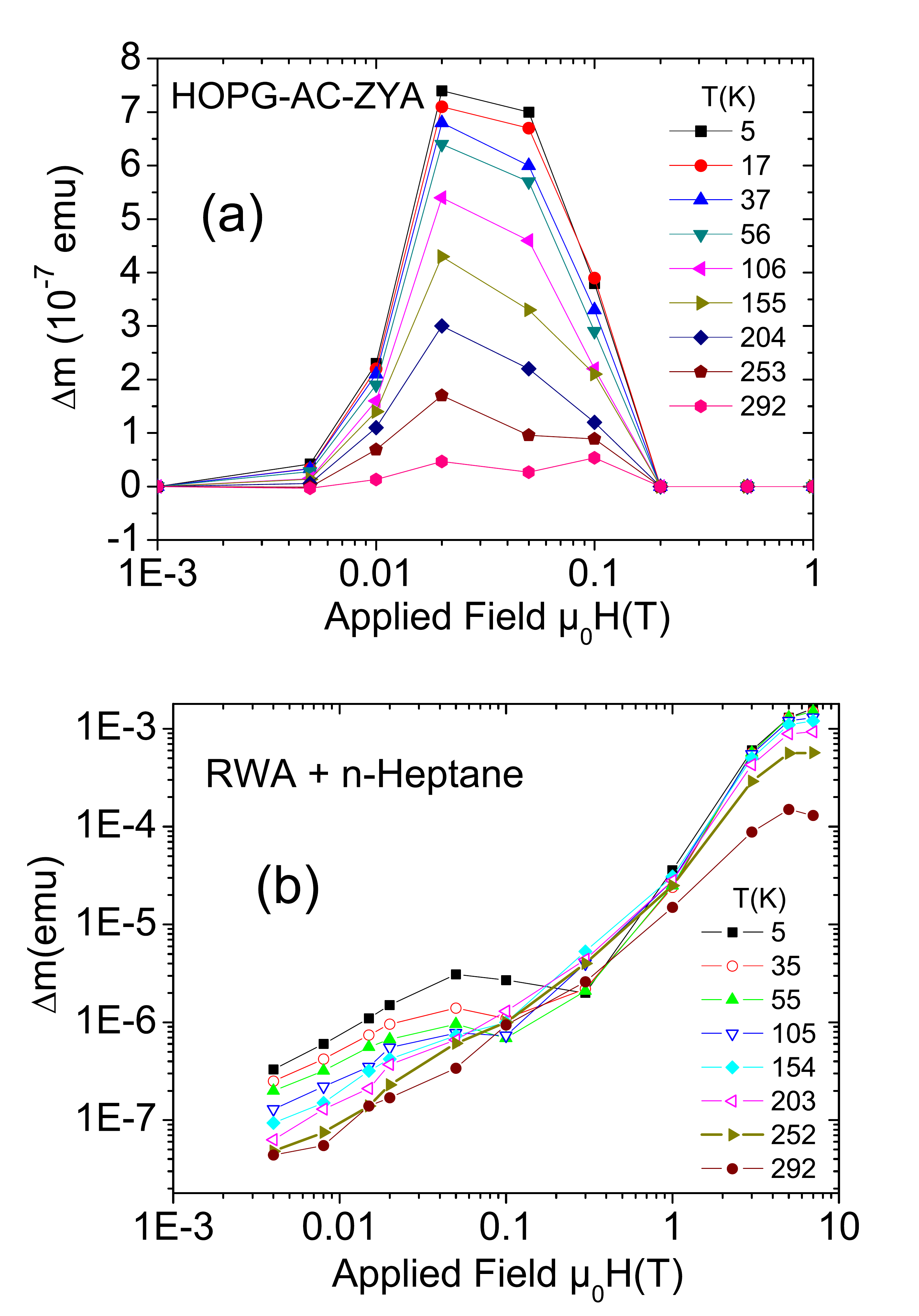}
\caption{The difference between FC and ZFC curves $\Delta m$ as a
function of applied field at different temperatures for the: (a)
bulk HOPG sample (field applied parallel to the graphene planes),
which magnetization shows a ferromagnetic hysteresis loop and (b)
a n-heptane-treated graphite powder, which shows a field
hysteresis similar to that of water treated graphite powder, see
Fig.~\ref{fig4}(b) and \cite{sch12}.}
\label{fig5}       
\end{figure}

In some cases the measurement of the intrinsic field hysteresis of
the superconducting regions is not straightforward to obtain
because of, either, a much larger magnetic background coming from
the rest of the sample,  or because the size of the
superconducting grains is too small and the effective pinning
strength for the magnetic entities is too weak to be measured
through a difference between ZFC and FC measurements. The
superconducting-like signals obtained in graphite samples from
magnetization measurements are mostly obtained after the necessary
subtraction of a large diamagnetic background \cite{yakovjltp00}.
Although this background is intrinsically non-hysteretic, its
contribution in real samples overwhelms the signals of interest by
up to two orders of magnitude, see, e.g.,
\cite{yakovjltp00,schcar}. In this case one may doubt whether the
subtraction of the diamagnetic slope is correctly done to obtain
results like in, e.g., Fig.~\ref{fig4}(b). In other words, if one
subtracts a tiny different diamagnetic slope from the original
data, in some cases one may transform the superconducting
hysteresis loops in ferromagnetic-like loops. Therefore, one needs
a background independent method that can rule out one of the two
possible origins for the hysteresis.

The thermomagnetic hysteresis (TH), i.e.,  the difference between
zero field cooling (ZFC) and field cooling (FC) curves at
different constant applied magnetic fields, is an important
experimental method to check for the existence of  pinning or
magnetic anisotropy of different kinds of  magnetic entities,
like, e.g., flux lines in superconductors as well as magnetic
domains in magnetically ordered materials. Thermomagnetic hysteresis is specially
useful to search for the existence of superconductivity in
materials where only a very small fraction of this phase exists,
like in granular superconductors.

In this work we studied the thermomagnetic response of different
graphite samples, in particular of ultra pure graphite powders
after n-heptane treatment. The study of the influence of n-heptane
on graphite flakes follows the results reported in
Ref.~\cite{kaw13}. We note, however, that the obtained results are
similar to graphite powders after water treatment. The main
results of this work is that the thermomagnetic hysteresis helps
to differentiate a ferromagnetic from a superconducting behavior,
supporting the superconducting interpretation of the field
hysteresis in \cite{yakovjltp00,sch12,schcar}.

The TH measurements of the magnetic moment $\Delta m(T) = m_{\rm
FC}(T) - m_{\rm ZFC}(T)$
 obtained from different graphite
powders and bulk samples were performed using a commercial MPMS-7
SQUID magnetometer. The temperature sweeps at fixed fields,
applied always at 5~K (ZFC state), were done with rates between 1
and 5 K/min. The differences in the observed hysteresis between
the rates lied within experimental error of $\Delta m(T) / m(T) <
0.3$\%. The used graphite powder was the same as previously
reported \cite{sch12}, i.e. RWA/T from SGL Carbon GmbH (Werk
Ringsdorff, Germany) with very low impurity concentration (e.g.,
Fe $<0.19~\mu$g/g). Further characterization of this graphite
powder and the influence of water on its magnetization was
thoroughly reported
 in \cite{sch12}.
In this study we found that if we follow the same procedure to
treat the graphite powder with n-heptane as was done with water,
i.e.  continuously stirring  of the graphite powder at room
temperature for 24~h   with 20~mL n-heptane
 (p.a. 99.99, Sigma-Aldrich) and
then filtered and dried at 100~C overnight, the change in the
magnetization was negligible. However, if the TH measurements of
the graphite powder started immediately after dropping a droplet
of n-heptane, the magnetization showed a behavior similar to the
water-treated graphite. All the powders were packed in polymer
foil (mass $<$ 10~mg). The magnetization of this foil was measured
independently and it gives a negligible magnetic moment in all the
here reported measurements in comparison with the measured powder
samples.

For comparison, we measured also: the untreated graphite powder
(RWA-virgin),  a pellet obtained from a high-purity graphite
cylinder used for spectroscopy  calibration (SK-AS01) and  a
graphite bulk sample (HOPG-AC-ZYA) from Advanced Ceramics, which
impurity concentration and magnetization behavior was throughly
characterized in \cite{spe14}. This last bulk sample shows a
ferromagnetic response
 for fields parallel to the graphene planes, originated by defects and/or hydrogen
with a saturation field of $\mu_0 H_{sat} \simeq 0.2~$T
\cite{spe14}. Similar field hysteresis curves at 300~K and the
corresponding XMCD hysteresis on the carbon K-edge can be seen in
\cite{ohldagnjp}.

In general the TH curve of a ferromagnetic sample as a function of
applied magnetic field starts at zero at zero field and tends to
vanish when the applied field is larger than the saturation field.
The reason for the behavior at large enough fields is easy to
understand. At the saturation field the sample is in one domain
state and no difference is measured between  $m_{\rm FC}(T)$ and
$m_{\rm ZFC}(T)$. The TH results for the graphite bulk sample with
ferromagnetic behavior are presented in Fig.~\ref{fig5}(a). We see
that the TH shows a maximum at $\sim 0.02~$T and vanishes at the
saturation field of 0.2~T. The overall hysteresis increases the
lower the temperature, as expected for magnetic graphite.

Similar TH measurements of the n-heptane treated graphite powder
with superconducting like field hysteresis (as in
\cite{sch12,schcar}) show a completely different behavior. The
influence of temperature on the TH curves is not monotonous in the
whole applied field, see the crossing of the curves at $\sim
0.3~$T in Fig.~\ref{fig5}(b). Moreover, the TH increases up to the
largest fields applied at temperatures below room temperature.
The decrease of the TH curves at  high enough fields and
temperatures can be interpreted as due to a weakening of the
pinning strength of the pinned entities, vortices and/or fluxons.
The high fields increase of the TH is not compatible with any
known magnetic order in carbon-based ferromagnetic materials or
other typical ferromagnets like magnetite (Fe$_3$O$_4$). On the
other hand, similar behavior is measured for granular Y123 high
temperature superconductor (unpublished). See also the studies
 in La$_{2-x}$Sr${_x}$CuO$_4$
 \cite{pan04,maj05}.  The crossing of the TH
curves at different temperatures at intermediate applied magnetic
fields is related to the granular
nature of the superconductivity in the sample.

Finally, magnetization measurements done on bulk highly oriented
pyrolytic graphite samples of high grade with interfaces, see
Section~\ref{inter}, show similar behavior as for the
water-treated graphite powder, in contrast to a bulk HOPG sample
without or less density of interfaces \cite{schcar}.
Several others experimental studies
partially reviewed in \cite{esqpip,chap7} indicate that superconductivity is embedded in
certain interfaces of the graphite structure. Structural evidence
for the existence of interfaces is presented in the next section.

\section{Direct evidence for the existence of interfaces in graphite}
\label{inter}

\begin{figure}
  \includegraphics[width=0.5\textwidth]{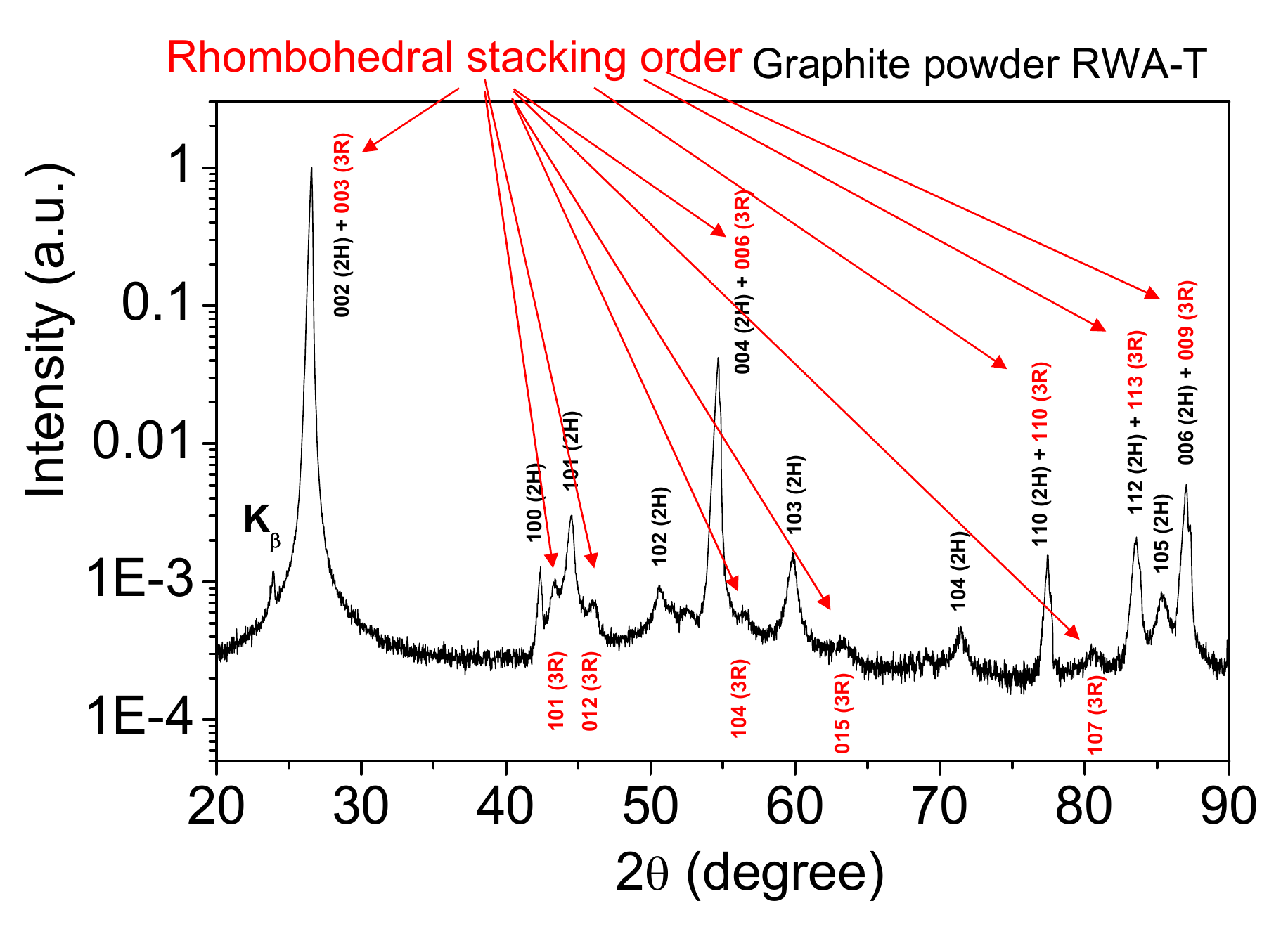}
\caption{X-rays diffraction pattern of an ultra pure graphite
powder (RWA-T) at room temperature. The labels near the Bragg
peaks indicate whether the maximum belong to Bernal (2H) or
rhombohedral (3R, in red color) phase. Some of the maxima coincide
with both phases within experimental resolution. Adapted from
\cite{pre16}.}
\label{XRD}       
\end{figure}

Several electrical resistance measurements of graphite samples of
different thickness published in the last years suggest that
graphite samples are not homogeneous and that the temperature as
well as the absolute value of the electrical resistivity and Hall
effect are not unique but thickness-
\cite{bar08,zor17,esq14,chap7} and sample-length dependent
\cite{dus11,esq11}. Whereas the last dependence is due to the
increasing contribution of ballistic transport to the total
resistance the smaller the sample length, the thickness dependence
observed in the transport properties is mainly due to the
existence of interfaces between the two stacking orders, hexagonal
or Bernal (2H) and rhombohedral (3R), see Fig.~\ref{XRD}, and
between two twinned regions around a common $c-$axis, see Fig.~5,
with the same or different stacking orders.

Figure~5 shows transmission electron microscope pictures
of regions in highly oriented pyrolytic graphite (HOPG) samples
(a-c) and of a natural graphite sample (d-f) obtained with the
electron bean parallel to the graphene layers. The different gray
colors mean a different electron diffraction due to a rotation of
the corresponding region respect to a common $c-$axis (always
normal to the graphene layers) or due to a different stacking
order. The 2D interfaces between those regions are very well
defined in well ordered samples (a,b,d,e) and less defined or
shorter in less ordered graphite samples (e.g. HOPG with grade B,
see Fig.~5(c)). The difficulty to pick up from a
macroscopic sample a region with a given density of interfaces is
clearly demonstrated in the pictures (d-f) where the density of
interfaces is not homogeneous in the same natural graphite sample.
All these pictures demonstrate the existence of interfaces and
the non-homogeneous nature of most of the graphite samples published in the
literature \cite{chap7}.

\begin{figure}
\label{TEM}       
\begin{center}
\includegraphics[width=0.50\textwidth]{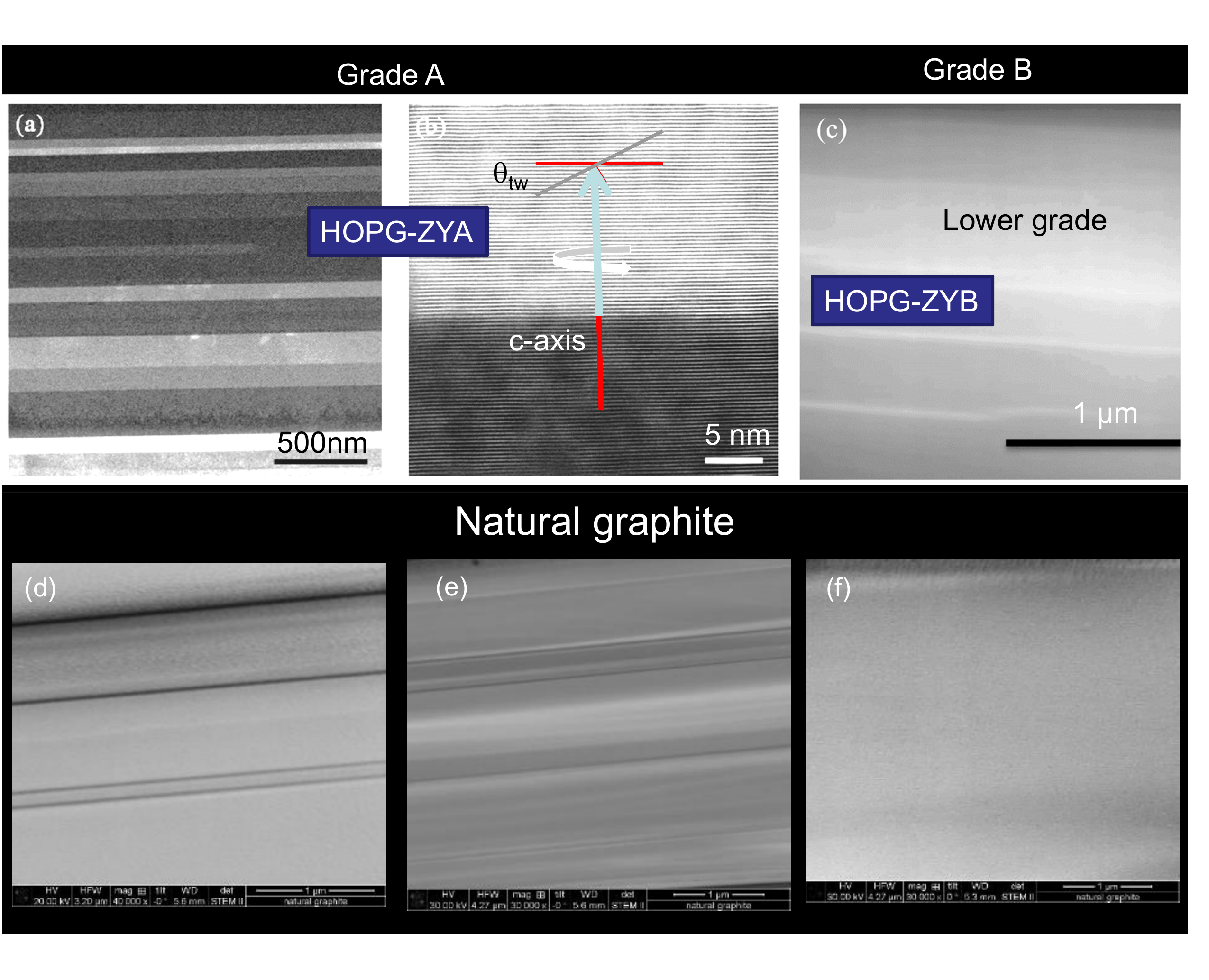}
\caption{(a-c): Transmission electron microscope pictures of
lamellae taken at low  (a) and high resolution (b) on highly
oriented pyrolytic graphite (HOPG) samples of grade A and at low
resolution on HOPG grade B (c). The pictures were obtained with
the electron beam parallel to the graphene layers. The sketch in
(b) indicates a possible rotation around the common $c-$axis of
the two crystalline regions identified through the  two different gray colors.
(c): The interfaces in this HOPG sample of lower grade are not so
well defined as in grade A (a-b) due to the disorder.
(d-f): Transmission electron microscope pictures on lamellae from the same
natural graphite sample but at different positions. The scale bar at the bottom right
indicates $1~\mu$m. Adapted from \cite{bar08,esqpip,esqarx14,pre16}.}
\end{center}
\end{figure}

From Fig.~5  it is clear that using graphite samples with
thickness smaller than the distance between interfaces, the larger
is the probability to measure an intrinsic property of graphite without a large
contribution from the interfaces,
as has been shown in several recent publications
\cite{bar08,zor17,esq14,chap7}. From all those experimental
studies we conclude that there are interfaces with metalliclike
properties. The metalliclike behavior of the transport properties
of graphite is not intrinsic of the graphite ideal structure
\cite{chap7}. Moreover, different theoretical works suggest that
some of the interfaces can have superconducting properties as,
e.g., between the 2H and 3R stacking orders
\cite{mun13,kop13,kopbook} as well as at the interfaces between
twinned regions with similar or different stacking orders
\cite{esqarx14}. According to those theoretical studies,
the main reason for the high temperature superconductivity is
related to possible flat bands \cite{kop11,vol13,hei16,kau16}
that can exist at certain interfaces \cite{fen12,col13,pie15} of the graphite structure.

\begin{figure*}
  \includegraphics[width=1\textwidth]{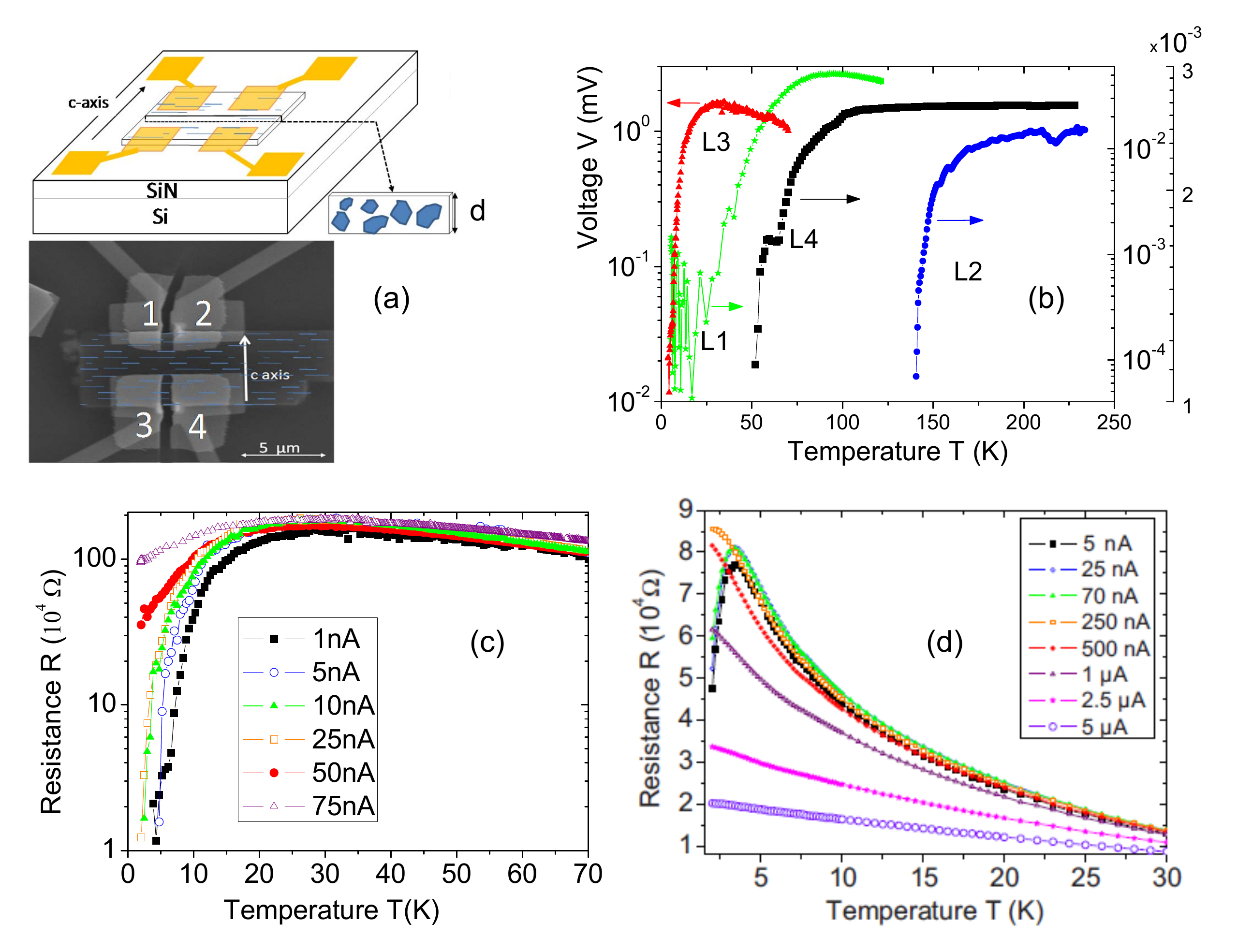}
\caption{(a): Sketch of a lamella on a dielectric substrate with
four contacts in a Van der Pauw-like configuration. The thickness
$d$ of the lamella can be between 100~nm and 800~nm. Note that the
direction of the $c-$axis of the graphite structure is parallel to
the substrate. The blue short lines indicate the interfaces with
probable superconducting regions, i.e. the blue regions in the
sketch at the right. Below the sketch we show a SEM picture of a
lamella cut from a bulk HOPG sample of grade A \cite{balthe}. (b)
Voltage vs. temperature measured at low input currents from four
different lamellae taken from the same HOPG sample of grade A. The
second $y-$axis corresponds to the lamella L4, adapted from
\cite{bal13}. (c) The resistance vs. temperature for the lamella
L3 at different input currents, adapted from \cite{bal13}. (d) As
(c) but for a lamella obtained from a HOPG sample of grade B, see
Fig.~5(c), adapted from \cite{bal15}. All measurements in this
figure were done at zero applied field.}
\label{fig6}       
\end{figure*}

\section{Contacting the interface edges in TEM lamellae}
\label{lame}

\begin{figure*}
  \includegraphics[width=1\textwidth]{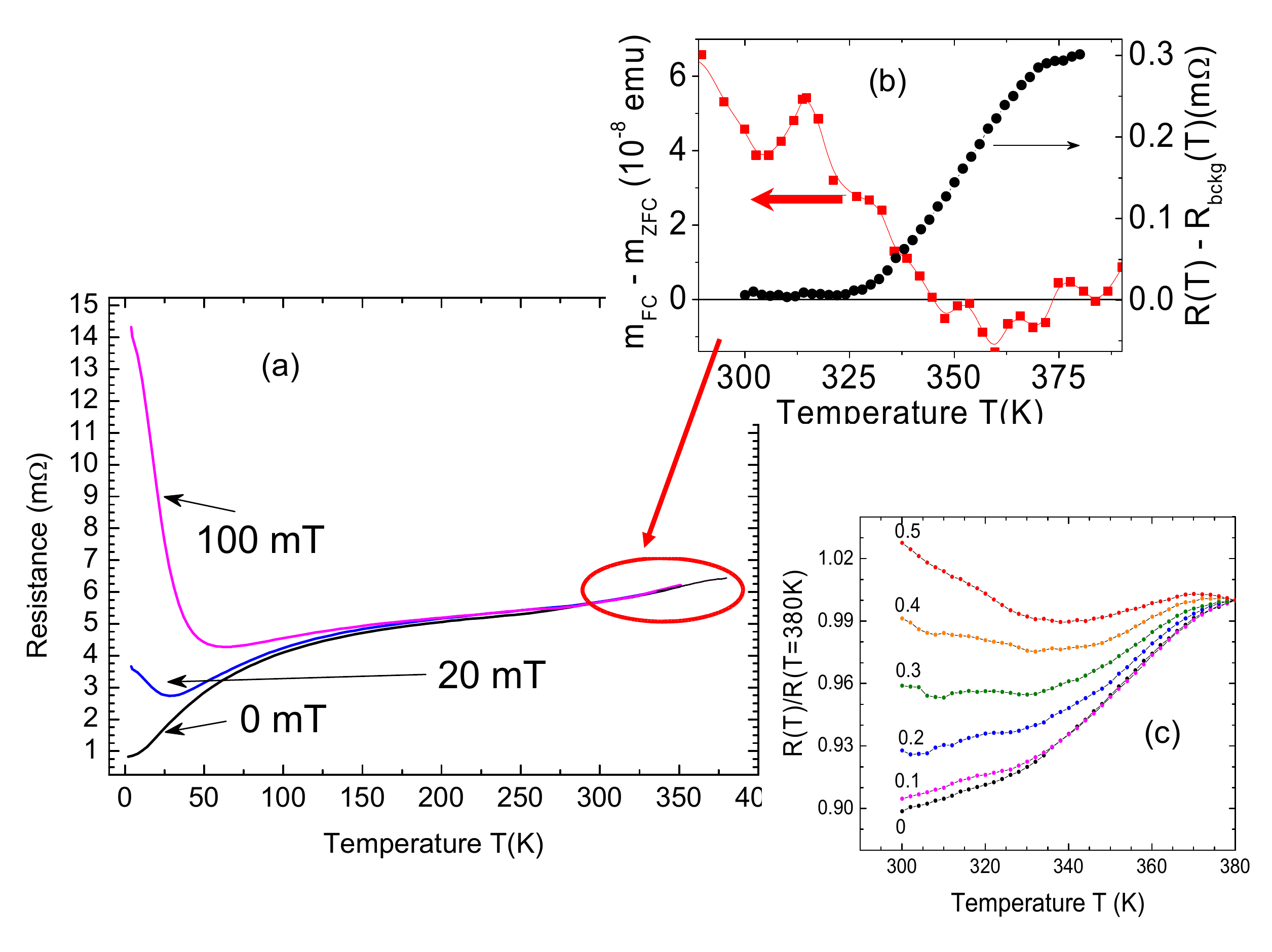}
\caption{Temperature dependence of the longitudinal electrical
resistance of a natural graphite sample from Brazil. (a): At
different magnetic fields applied normal to the graphene planes
and interfaces.  (b): Temperature dependence of the difference
between FC  and ZFC
 magnetic moment (left $y-$axis) measured with
a SQUID, after applying a field of 50~mT at 250~K with the sample
in the virgin state. Right $y-$axis: Temperature dependence of the
resistance at zero field from (c), after subtracting a straight
line background to show the clear change in slope of the measured
resistance, indicating a well-defined transition region.  (c):
Normalized resistance data at high temperatures. The numbers at
the curves indicate the field (in Tesla) applied at 380~K and the
measurements were done decreasing temperature. The natural
graphite sample had 0.4~mm length between voltage electrodes and
$\simeq 1~$mm width. Adapted from \cite{pre16}.}
\label{fig7}       
\end{figure*}

Although signs of granular superconductivity in the electrical resistance of
graphite samples with electrodes on the main surface (usually on the top  graphene layer)
have been  recognized in earlier studies \cite{esq08,sru11}, if we could contact
directly the edges of the interfaces found in graphite
samples we expect to observe more clear signs of Josephson or granular superconducting behavior in the
transport measurements. As it became evident from the results shown in Fig.~5,
added to the unknown position of the possible interfaces with superconducting regions in
a given graphite bulk sample,
this kind of experimental work is highly time consuming, and was done based  mainly on a trial and error
strategy. Note that one prepares a TEM lamella of a few micrometers area parallel to the
$c-$axis, see Fig.~\ref{fig6}(a), from a graphite sample of mm$^2$ area.  Additionally,
good electrical contacts on the edges of the interfaces are difficult to prepare after cutting
the graphite sample using FIB and Ga$^+$ ions.

Several results of different TEM lamellae were published in
\cite{bal13,bal15,bal14I}. Selected results are included in Fig.~\ref{fig6}(b-d). The main messages from
these results can be summarized as follows: - The voltage of the TEM samples measured at  low input
currents show a clear drop below a sample dependent temperature, see Fig.~\ref{fig6}(b). - The
voltage or resistance curves vs. temperature depend strongly on the input current, see Fig.~\ref{fig6}(c,d).
- The more disordered the graphite structure or the smaller the size or length of the interfaces
the lower is the temperature where superconducting-like  behavior is observed \cite{bar15,bal14I}. - Finally, the
$I-V$ curves measured at different temperatures \cite{bal13} are compatible with the behavior expected
for Josephson junctions under the influence of thermal activation \cite{amb69}. This means that linear
$I-V$ curves are expected for Josephson junctions at high enough temperatures, as has been measured
in high-temperature superconductors at high enough temperatures \cite{gro90}. In other words, one
could have a sample where due to the size of the superconducting regions and their Josephson coupling the
resistance at high temperatures remains finite and nearly ohmic, making difficult to recognize from these measurements
alone the existence of
superconductivity.

The transport results obtained from TEM lamellae indicate the influence of
the Josephson effect on the transport properties, an influence that increases the
lower the temperature, where the
Josephson coupling and/or the size of the superconducting regions are large
enough. The clear decrease in the resistance or voltage at constant current depicted in
Fig.~\ref{fig6}(b) can be interpreted as due to an enhancement of the superconducting
links between the superconducting regions existing at certain interfaces. In other words,
those clear drops in the voltage do not indicate necessarily the critical temperature
of those regions. The results discussed in the next section provide a clear evidence for a
critical temperature of the superconducting regions above room temperature.

\begin{figure*}
  \includegraphics[width=1\textwidth]{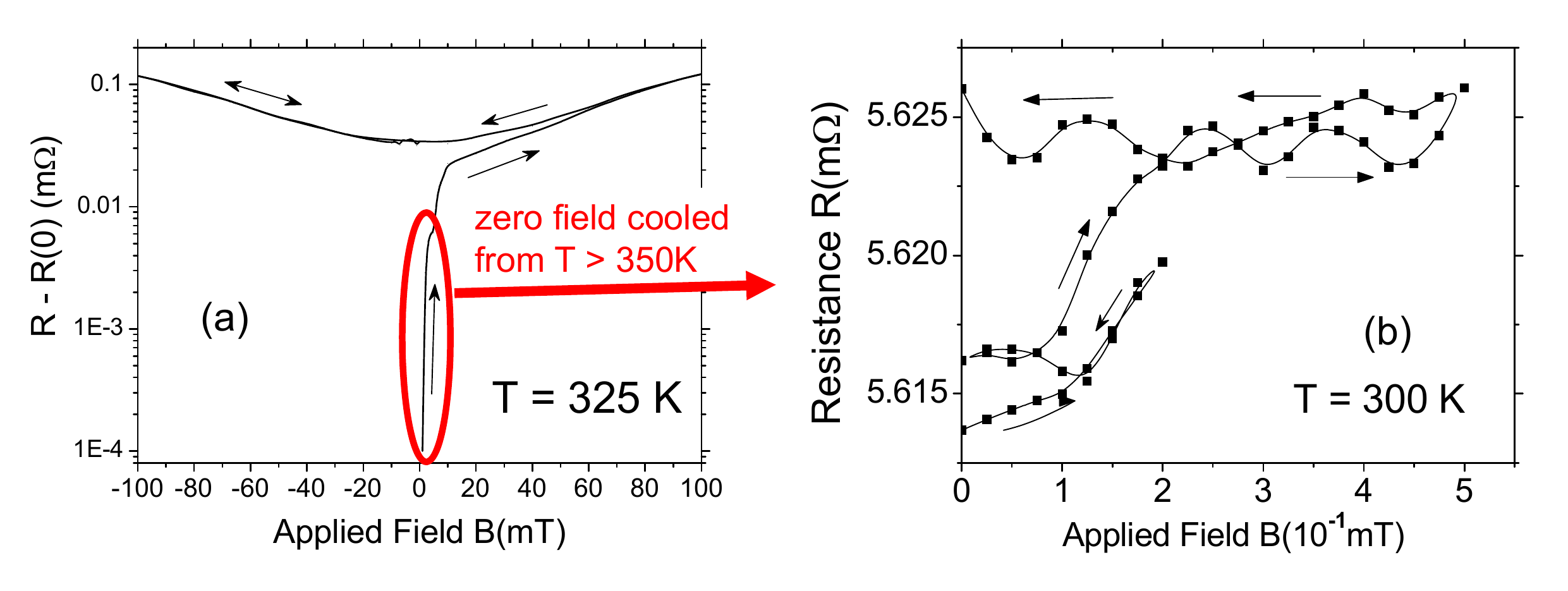}
\caption{(a) Difference between the resistance and its value at
zero field vs. applied field in two opposite directions at 325~K
measured in a natural graphite sample. The arrows indicate the
sweep field direction. The virgin state is reached after zero
field cooling the sample from temperature above 350~K. (b) Similar
as in (a) but at 300~K and at very low fields  starting from the
virgin state after ZFC from 390~K. Note the clear remanence after
decreasing the magnetic field to zero. Adapted from \cite{pre16}.}
\label{fig8}       
\end{figure*}

\section{Identifying the superconducting critical temperature}
\label{Tc}

Following the arguments developed in the last section,  if the results of magnetization
obtained at 300~K, see Fig.~\ref{fig4}(b) and \cite{schcar},  as well as the results of Fig.~\ref{fig1},
are an indication
of superconductivity, we then should
expect a critical temperature clearly above room temperature. To identify it, high-resolution
electrical resistance measurements on different natural graphite samples as well as HOPG samples
were reported recently \cite{pre16}. The main results obtained in a natural graphite sample from a
Brazil mine are shown in Figs.~\ref{fig7} and \ref{fig8}. The resistance of the sample was measured with  four electrodes
at the top surface. The temperature dependence of the resistance for three different magnetic fields applied normal
to the graphene and interface planes is shown in Fig.~\ref{fig7}(a). The huge magnetic field driven
metal-insulator transition (MIT) at $T < 150~$K is especially large for the measured natural graphite sample. This
 field driven MIT in graphite was speculated in the past to be  related to superconductivity \cite{kempa00,yakovadv03}.
Nowadays we know that the MIT in graphite as well as the metalliclike behavior of its resistance
are not intrinsic but are due to the contribution of certain interfaces in parallel to the graphene layers
of the graphite structure \cite{bar08,zor17,chap7}.

If one measures the resistance with care, see \cite{pre16} for more details, one can identify an anomaly
at $T \sim 350~$K for that sample. After subtracting a linear in temperature background to the measured
resistance we obtain the curve in  Fig.~\ref{fig7}(b), which clearly shows a transition-like behavior.
The TH at a field of 50~mT is shown in the same figure, which result indicates the start of
an irreversibility  at a similar temperature as the transition in the resistance. Figure~\ref{fig7}(c) shows the
temperature dependence of the normalized resistance at different applied fields. The magnetoresistance results also indicate
the existence of an anomaly at such high temperatures.

A  support for the interpretation of a superconducting transition at such high temperatures is
given by the large irreversibility and remanence measured in the resistance \cite{pre16}. As an
example, we show in Fig.~\ref{fig8} the irreversibility in the magnetoresistance at $T < T_c$ after
ZFC from $T > 350~$K. The fact that no magnetoresistance is measurable in this field range or higher within a
relative resolution
better than $10^{-5}$ when the field is applied parallel to the interfaces, rules out a relation
of the transition to a magnetic order phenomenon.

Because the superconducting regions are thought to be within certain interfaces,
there is no simple experimental method to demonstrate zero electrical resistance
below $T_c$, unless one tries to contact directly the superconducting regions, which location
within the sample and within a given interface remains unknown. Moreover, if the size of the superconducting
regions is much smaller than the effective London penetration depth, in addition to
demagnetization effects, the flux expulsion, i.e. the Meissner effect, would be practically immeasurable.
An alternative proof for the existence of superconductivity, i.e. of
a region in the sample with real zero resistance, can rely on the observation of dissipationless currents
that maintain a magnetic flux trapped at certain regions
of the sample. Using magnetic force microscopy (MFM) we were able to
localize a permanent current path that was triggered in a natural graphite bulk sample after removing
an applied magnetic field, i.e. at remanence \cite{sti17}, as expected from the results
shown in Fig.~\ref{fig8}.

\begin{figure*}
  \includegraphics[width=1\textwidth]{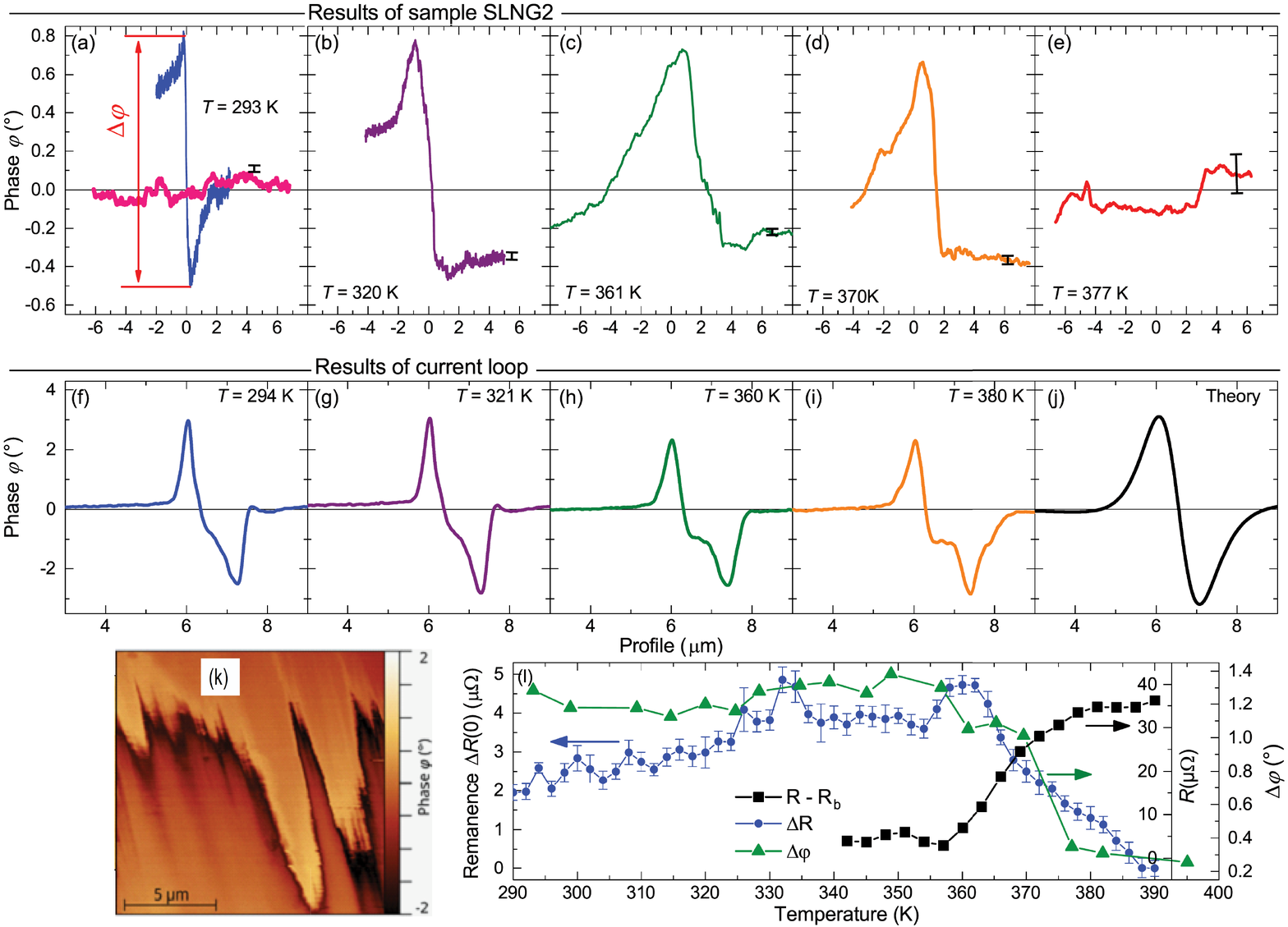}
\caption{(a)--(e): Line scans of the field gradient at different temperatures
  obtained at the edges of the trapped flux region in remanence.
  The error bars represent the standard deviation of the
phase at the given temperature.  The blue curve in (a) is the measured line scan with the sample
in remanent state and the purpure color curve is the line scan at the same position but cooling the sample
at zero field after reaching 395~K and the magnetic flux vanishes, see (l).
  (f-i): The same but  for a current
  loop of Au of ring geometry and with $\approx 1~\mu$m
  width prepared by electron lithography.
A theoretical phase shift is shown in (j), obtained
  from  a simulated current line loop of  ring geometry and zero width. The phase
  image  (k)  shows a small part of the whole current loop in a region of the sample surface where the current
  path is measured (the dark edge between  the regions of different
  phase colors).
 The meandering structure of the current path shown by the MFM phase
line looks similar to the one
observed in high-temperature superconducting oxides in
remanence~\cite{vla97,vla98}.
  The line scans of the phase (a-e) were taken normal to
  the current line.  In Fig.(l), the phase difference $\Delta\varphi$ (see
  (a)), the resistance $R$ (after a linear background subtraction)
  and the remanence $\Delta R(0)$ (see \cite{pre16} for details on the
  remanence measurements) are shown as
  function of temperature. Adapted from \cite{sti17}.}
\label{fig9}       
\end{figure*}

Figure~\ref{fig9}(a) shows the line scans of the phase measured with a MFM tip and obtained across
the current line region, see Fig,~\ref{fig9}(k), measured at 293~K. The measured curve is compatible with
the existence of a electrical current
path, as the measurements on a Au current loop, Figs~\ref{fig9}(h-i) and the theoretical
line scan in Fig.~\ref{fig9}(j) indicate. The amplitude of the
jump in the line scan $\Delta \varphi$, defined in Fig.~\ref{fig9}(a), is proportional to the
current amplitude. The temperature dependence of this phase jump shown in
Fig.~\ref{fig9}(l) vanishes irreversibly at the same transition temperature obtained by
the electrical resistance  and  the remanence for the same sample, for more details see \cite{sti17}.

\section{Conclusion}
\label{Con}

The experimental facts obtained the last  years provide evidence for 2D
granular superconductivity at certain interfaces in graphite.
High resolution transport and magnetization measurements, as well as MFM
indicate the existence of a transition at a critical temperature $T_c \gtrsim 350~$K. The remanence in the electrical resistance
as well as the phase signals obtained from MFM provide evidence  for persistent currents up to $T_c$, evidence
that indicate the existence of regions with negligible electrical resistance.
On the other hand there is a  ``critical temperature" $T_c^J$, below which the  Josephson coupling
between grains is large enough to strongly
influence the electrical resistance of graphite samples with interfaces. This   $T_c^J$
depends on the interface size and/or internal order in the graphite structure.
There are several open questions to be answered in the future. Namely
how large are the critical fields, are protons (or hydrogen) playing any role at the interfaces,
and how to produce specific interfaces? A real help for experimentalists is the possibility
to use scanning magnetic
imaging techniques  to identify the regions of graphite
samples were superconductivity is localized. This will undoubtedly
help to answer some of the open questions and further characterize the superconducting regions
in graphite. If the production of specific graphite interfaces
turns out to be very difficult, the localization of the superconducting regions would
already  pave the way for
future device implementations of graphite mine flakes, which costs nowadays are much less than
an artificial production.

\begin{acknowledgements}
We thank Ana Champi and Henning Beth for discussions and support and for providing us with the natural graphite samples.
We thank T. Heikkil\"a, G. Volovik, and J. Meijer for fruitful discussions.
 C.E.P. gratefully acknowledges
 the support provided by The Brazilian National Council for the Improvement of
 Higher Education (CAPES). M.S. and J.B-Q. are supported by the DFG
 collaboration project SFB762.
\end{acknowledgements}

\bibliographystyle{spphys}       

\end{document}